\documentclass[preprint,preprintnumbers,a4paper,nofootinbib]{revtex4-1}

\usepackage{amsmath,amssymb}
\usepackage[utf8]{inputenc}
\usepackage{color}
\usepackage{graphicx}
 \usepackage{verbatim} 
\allowdisplaybreaks[1]

\newcommand{\tr}{\mathop{\mathrm{tr}}}

\begin{document}

\title{Tensor network representation of non-abelian gauge theory coupled to reduced staggered fermions}

\preprint{}

\author{Muhammad Asaduzzaman}
\email[]{muhammad-asaduzzaman@uiowa.edu}
\affiliation{Department of Physics and Astronomy, The University of Iowa, Iowa City, Iowa 52242, USA}

\author{Simon Catterall}
\email[]{smcatter@syr.edu}
\affiliation{Department of Physics, Syracuse University, Syracuse, NY 13244, USA}

\author{Yannick Meurice}
\email[]{yannick-meurice@uiowa.edu}
\affiliation{Department of Physics and Astronomy, The University of Iowa, Iowa City, Iowa 52242, USA}

\author{Ryo Sakai}
\email[]{r.sakai@j-ij.com}
\affiliation{\textit{Jij Inc.}, Bunkyo-ku, Tokyo 113-0031, Japan}
\affiliation{Department of Physics, Syracuse University, Syracuse, NY 13244, USA}

\author{Goksu Can Toga}
\email[]{gctoga@syr.edu}
\affiliation{Department of Physics, Syracuse University, Syracuse, NY 13244, USA}

\date{\today}

\begin{abstract}
  We show how to construct a tensor network representation of the path integral
  for reduced staggered fermions coupled to a non-abelian
  gauge field in two dimensions. The resulting
  formulation is both memory and computation efficient because reduced staggered fermions can be represented in terms of a minimal number of tensor
  indices while the gauge sector can be approximated using Gaussian quadrature with a truncation.
  Numerical results obtained using the Grassmann TRG
  algorithm are shown for the case of $SU(2)$ lattice gauge theory
  and compared to Monte Carlo results.
\end{abstract}

\maketitle

\section{Introduction}

Tensor networks furnish a powerful tool to represent and study lattice quantum field theories.
In a Hamiltonian formulation they yield efficient representations
of low lying states of the system \cite{Orus:2013kga,Banuls:2018jag}
while in the context of a Euclidean path integral they form the starting point of efficient
blocking/RG schemes that can be used to compute a variety of observable.

One of the main motivations for their use
within the HEP community
is the famous sign problem that prohibits the use of Monte Carlo
techniques for many theories of interest.
In contrast, renormalization group algorithms for tensor
networks are deterministic and hence insensitive to sign problems---see~\cite{Meurice:2022xbk,Meurice:2020pxc} for reviews
and recent developments.

The ultimate goal in HEP is to formulate a tensor network
representation of full QCD, in which fermions are coupled to an $SU(3)$ gauge field in four dimensions which can be contracted efficiently on current hardware.\footnote{By taking the time
  continuum limit one can also extract a gauge invariant Hamiltonian from such a network that
  can be implemented, in principle, on quantum computers.}

The numerical complexity, in terms of both CPU and memory, of any tensor network depends on the number of physical degrees of freedom which must be captured in the tensor. For the gauge
fields one must truncate the continuous degrees of freedom associated with the gauge group
down to a finite set while
fermions are characterized by multidimensional bond dimensions (see \textit{e.g.} \cite{Gu:2010yh,Gu:2013gba}). In addition the number of
tensor indices increases rapidly with dimension. These facts imply that
tensor renormalization group computations for the simplest non-abelian
lattice gauge theory coupled to fermions are already extremely
difficult even in two space time dimensions~\footnote{
  Two dimensional QCD was studied using tensor networks in ref.~\cite{Bloch:2022vqz}.
  In that paper the strong coupling limit is taken, so that the major part of the physical degrees of freedom are integrated out at the initial stage.
  By contrast, our current paper provides a way to construct
  a tensor network representation for QCD-like theories for any value of the coupling constant.
}~\footnote{
Note that theories where $SU(2)$ gauge fields are coupled to scalar fields have been studied in refs.~\cite{Bazavov:2019qih,Asaduzzaman:2019mtx}.
}.

A typical way to extract discrete tensor indices for gauge or spin systems
is the character expansion and this approach has been shown to be
successful for studies of $U(N)$ and pure $SU(N)$ LGTs~\cite{Shimizu:2014uva,Liu:2013nsa,Hirasawa:2021qvh}.
Recently other approaches that are based on the
method of quadratures, probabilistic sampling, and trial (variational) actions have been proposed~\cite{Kuramashi:2019cgs,Fukuma:2021cni,Kuwahara:2022ubg}~\footnote{
Note that the use of the quadrature method was introduced earlier in the context of scalar fields~\cite{Kadoh:2018hqq,Kadoh:2018tis}.}.
Also a new method in which the tensors depend on only representation
indices was proposed in~\cite{Yosprakob:2023jgl} for pure gauge theories.

In this work, we discretize the path integral using the Gaussian quadrature rule. Since the fermions are represented by Grassmann valued fields they are naturally discrete.
Nevertheless the requirements needed to build Grassmann tensor networks are typically large
since they depend on the number of both spinor and color components of a complex field. Using ordinary staggered fermions removes the spinor index component but we will show that
it still leaves a formidable computational challenge even in the simplest case of a two
color gauge theory. In contrast,
we will show that
{\it reduced} staggered fermions \cite{vandenDoel:1983mf} give the most economical lattice fermion formulation possible
in such systems. Reduced staggered fermions are also interesting in the context of
symmetric mass generation and recent
efforts to construct chiral lattice gauge theories---see \cite{Catterall:2022jky,Butt:2021koj}.
Indeed in the latter case a sign problem is almost inevitable which provides strong motivation
for the use of tensor methods.

\section{Model and tensor network representation}

As a warmup we will focus first on the construction of a theory of regular staggered fermions coupled to $SU(2)$---the simplest continuous non-abelian gauge group. First, we describe why this theory is computationally challenging in the tensor renormalization group studies. Subsequently we introduce a tensor network formulation for the $SU(2)$ gauge theory with reduced staggered fermions where the higher order orthogonal iteration (HOOI) algorithm is used for the construction of tensor.

\subsection{$SU(2)$ theory with full staggered fermions}

We can make a tensor network representation of this fermion model
by following the Grassmannn tensor network construction
(see \textit{e.g.} \cite{Takeda:2014vwa}). First we express the action as a product of Grassmannn valued tensors.
The action for the gauged staggered fermion is given by
\begin{align}
  S_{\mathrm{F}}\left[ U \right]
  = \sum_{n}
  \left[
  m \bar{\psi}_{n}\psi_{n}
  + \sum_{\mu=1}^{2} \frac{\eta_{n,\mu}}{2} \left( \bar{\psi}_{n}U_{n,\mu}\psi_{n+\hat{\mu}} - \bar{\psi}_{n+\hat{\mu}}U_{n,\mu}^{\dagger}\psi_{n} \right)
  \right].
\end{align}
The staggered sign factor is defined by $\eta_{n,\mu} = (-1)^{\sum_{\nu<\mu}n_{\nu}}$. Both periodic and anti-periodic boundary conditions can be used.

The partition function can be expanded thanks to the nilpotency of the Grassmannn variables:
\begin{align}
  Z_{\mathrm{F}}\left[ U \right]
  & = \int \mathcal{D}\bar{\psi}\mathcal{D}\psi
    \prod_{n} e^{-S_{\mathrm{F}}[U]} \nonumber \\
  & =
    \begin{aligned}[t]
      \int \mathcal{D}\bar{\psi}\mathcal{D}\psi
      \prod_{n} & \prod_{a=1}^{2} \sum_{s_{n}^{a}=0}^{1} \left( -m\bar{\psi}_{n}^{a}\psi_{n}^{a} \right)^{s_{n}^{a}} \\
                & \cdot \prod_{a,b=1}^{2}
                  \begin{aligned}[t]
                    & \sum_{x_{n,1}=0}^{1} \left( -\frac{\eta_{n,1}}{2} \bar{\psi}_{n}^{a}U_{n,1}^{ab}\psi_{n+\hat{1}}^{b} \right)^{x_{n,1}^{ab}}
                      \sum_{x_{n,2}=0}^{1} \left( \frac{\eta_{n,1}}{2} \bar{\psi}_{n+\hat{1}}^{a}U_{n,1}^{ba\ast}\psi_{n}^{b} \right)^{x_{n,2}^{ab}} \\
                    & \cdot \sum_{t_{n,1}=0}^{1} \left( -\frac{\eta_{n,2}}{2} \bar{\psi}_{n}^{a}U_{n,2}^{ab}\psi_{n+\hat{2}}^{b} \right)^{t_{n,1}^{ab}}
                      \sum_{t_{n,2}=0}^{1} \left( \frac{\eta_{n,2}}{2} \bar{\psi}_{n+\hat{2}}^{a}U_{n,2}^{ba\ast}\psi_{n}^{b} \right)^{t_{n,2}^{ab}}.
                  \end{aligned}
    \end{aligned}
\end{align}
As shown in \cite{Takeda:2014vwa}, the lattice coordinates $x$ and $t$ which label the index associated with the expansion of the exponential constitute candidates for the tensor indices.
On each link, and for
both $\psi$ and $\bar{\psi}$, there is a two component (forward and backward hopping)
index and, in addition, a color index running over two values for $SU(2)$.
Thus, the bond dimension associated with each
fermion link will turn out to be $2^{2 \times 2 \times 2} = 256$.
This is prohibitively large since, in the complete tensor network, one has to consider additionally the contribution from the gauge part.
Specifically, if we assume that the bond dimension of the gauge sector is $\chi$, the bond dimension of the total tensor network will be $256\chi$, and this is not currently feasible~\footnote{
  In previous tensor network studies, the typical bond dimension is $100$ or less.
  While bond dimensions as large as 512 have been used for
  the two dimensional Ising model~\cite{2018PhRvE..97c3310M}, such bond dimensions require  a huge amount
  of CPU time
  and also carry memory footprints on the order of 100--1000 GB.
}.
To remedy this situation we have instead considered using reduced staggered fermions.

\subsection{$SU(2)$ theory with reduced staggered fermions}

If one uses a massless reduced staggered formulation as in ref.~\cite{Catterall:2018pms}, the degrees of freedom can be reduced by half. We substitute the staggered fields by the reduced staggered fermions using the transformation $\psi_n \to (1-\epsilon_n)\psi_{n}/2$ and $\bar{\psi}_n \to (1+\epsilon_n)\psi_{n}/2$. In this formulation the reduced staggered field $\psi_n$  and it's conjugate $\bar{\psi}_n$ are placed on odd and even sites (or even and odd sites), respectively, so that one can just relabel $\bar{\psi_n}$ as $\psi_n^{\mathrm{T}}$. The fermionic action can  then be simplifed to
\begin{align}
  S_{\mathrm{F}}\left[ U \right] = \sum_{n} \sum_{\mu=1}^{2} \frac{\eta_{n,\mu}}{2} \psi_{n}^{\mathrm{T}} \mathcal{U}_{n,\mu} \psi_{n+\hat{\mu}}.
\end{align}
A ``projected'' link variable $\mathcal{U}$ is defined by $\mathcal{U} = (1+\epsilon_{n})U_{n,\mu}/2 + (1-\epsilon_{n})U_{n,\mu}^{\ast}/2$, where the parity factor is $\epsilon_{n} = (-1)^{n_{1}+n_{2}}$.

In this case the Boltzmann factor is expanded like
\begin{align}
  e^{-S_{\mathrm{F}}}
  & = \sum_{\left\{ x,t \right\}} \prod_{n} \prod_{a,b=1}^{2}
    \left( - \frac{\eta_{n,1}}{2} \psi_{n}^{a} \mathcal{U}_{n,1}^{ab} \psi_{n+\hat{1}}^{b} \right)^{x_{n}^{ab}}
    \left( - \frac{\eta_{n,2}}{2} \psi_{n}^{a} \mathcal{U}_{n,2}^{ab} \psi_{n+\hat{2}}^{b} \right)^{t_{n}^{ab}}.
\end{align}

Because of the halving of degrees of freedom the bond dimension of the resultant fermion tensor network is now just $2^{2\times 2}=16$.
This is a significant reduction from a bond dimension of
$256$ for the case of full staggered fermions.

We can split $\psi_{n}^{a}\psi_{n+\hat{1}}^{b}$ and $\psi_{n}^{a}\psi_{n+\hat{2}}^{b}$ using a set of dummy Grassmannn variables $\alpha_n, \beta_n$ as
\begin{align}
  & \psi_{n}^{a}\psi_{n+\hat{1}}^{b} = \int (\psi_{n}^{a}\mathrm{d}\alpha_{n}^{ab}) (\mathrm{d}\bar{\alpha}_{n+\hat{1}}^{ab}\psi_{n+\hat{1}}^{b}) (\bar{\alpha}_{n+\hat{1}}^{ab}\alpha_{n}^{ab}), \nonumber \\
  & \psi_{n}^{a}\psi_{n+\hat{2}}^{b} = \int (\psi_{n}^{a}\mathrm{d}\beta_{n}^{ab}) (\mathrm{d}\bar{\beta}_{n+\hat{2}}^{ab}\psi_{n+\hat{2}}^{b}) (\bar{\beta}_{n+\hat{2}}^{ab}\beta_{n}^{ab}).
\end{align}

Using dummy Grassmannn variables, the Boltzmann factor turns out to be
\begin{align}
  e^{-S_{\mathrm{F}}} =
  \sum_{\left\{ x,t \right\}} \prod_{n} \prod_{a,b=1}^{2}
  & \left( \frac{\eta_{n,1}}{2} \mathcal{U}_{n,1}^{ab} \right)^{x_{n}^{ab}}
    \left( \frac{\eta_{n,2}}{2} \mathcal{U}_{n,2}^{ab} \right)^{t_{n}^{ab}} \nonumber \\
  & \cdot (\psi_{n}^{a}\mathrm{d}\alpha_{n}^{ab})^{x_{n}^{ab}}
    (\psi_{n+\hat{1}}^{b}\mathrm{d}\bar{\alpha}_{n+\hat{1}}^{ab})^{x_{n}^{ab}}
    (\psi_{n}^{a}\mathrm{d}\beta_{n}^{ab})^{t_{n}^{ab}}
    (\psi_{n+\hat{2}}^{b}\mathrm{d}\bar{\beta}_{n+\hat{2}}^{ab})^{t_{n}^{ab}} \nonumber \\
  & \cdot (\bar{\alpha}_{n+\hat{1}}^{ab}\alpha_{n}^{ab})^{x_{n}^{ab}} (\bar{\beta}_{n+\hat{2}}^{ab}\beta_{n}^{ab})^{t_{n}^{ab}}.
\end{align}

Then the fermion partition function can  be expressed as
\begin{align}
  Z_{\mathrm{F}}\left[ U \right] & = \int \left( \prod_{n} \mathrm{d}\psi_{n}^{1} \mathrm{d}\psi_{n}^{2} \right) e^{-S_{\mathrm{F}}} \nonumber \\
                                 & = \sum_{\{x,t\}} \prod_{n} \left[ \prod_{a,b=1}^{2} \left( \mathcal{U}_{n,1}^{ab} \right)^{x_{n}^{ab}} \left( \mathcal{U}_{n,2}^{ab} \right)^{t_{n}^{ab}} \right]T_{\mathrm{F} x_{n}t_{n}x_{n-\hat{1}}t_{n-\hat{2}}} G_{n, x_{n}t_{n}x_{n-\hat{1}}t_{n-\hat{2}}},
\end{align}
where, the bosonic and the fermionic components can be written repectively as
\begin{align}
  T_{\mathrm{F} x_{n}t_{n}x_{n-\hat{1}}t_{n-\hat{2}}}
  = \int \mathrm{d}\psi_{n}^{1} \mathrm{d}\psi_{n}^{2}
  & \left[ \prod_{a,b=1}^{2} \left( \frac{\eta_{n,1}}{2} \right)^{x_{n}^{ab}} \left( \frac{\eta_{n,2}}{2} \right)^{t_{n}^{ab}} \right] \nonumber \\
  & (\psi_{n}^{2})^{t_{n-\hat{2}}^{22}} (\psi_{n}^{2})^{t_{n-\hat{2}}^{12}} (\psi_{n}^{1})^{t_{n-\hat{2}}^{21}} (\psi_{n}^{1})^{t_{n-\hat{2}}^{11}} \nonumber \\
  & \cdot (\psi_{n}^{2})^{x_{n-\hat{1}}^{22}} (\psi_{n}^{2})^{x_{n-\hat{1}}^{12}} (\psi_{n}^{1})^{x_{n-\hat{1}}^{21}} (\psi_{n}^{1})^{x_{n-\hat{1}}^{11}} \nonumber \\
  & \cdot (\psi_{n}^{2})^{t_{n}^{22}} (\psi_{n}^{1})^{t_{n}^{12}} (\psi_{n}^{2})^{t_{n}^{21}} (\psi_{n}^{1})^{t_{n}^{11}} \nonumber \\
  & \cdot (\psi_{n}^{2})^{x_{n}^{22}} (\psi_{n}^{1})^{x_{n}^{12}} (\psi_{n}^{2})^{x_{n}^{21}} (\psi_{n}^{1})^{x_{n}^{11}},
\end{align}
and
\begin{align}
  G_{n, ijkl} =
  & \left( \mathrm{d}\alpha_{n}^{11} \right)^{x_{n}^{11}} \left( \mathrm{d}\alpha_{n}^{21} \right)^{x_{n}^{21}} \left( \mathrm{d}\alpha_{n}^{12} \right)^{x_{n}^{12}} \left( \mathrm{d}\alpha_{n}^{22} \right)^{x_{n}^{22}} \nonumber \\
  & \cdot \left( \mathrm{d}\beta_{n}^{11} \right)^{t_{n}^{11}} \left( \mathrm{d}\beta_{n}^{21} \right)^{t_{n}^{21}} \left( \mathrm{d}\beta_{n}^{12} \right)^{t_{n}^{12}} \left( \mathrm{d}\beta_{n}^{22} \right)^{t_{n}^{22}} \nonumber \\
  & \cdot \left( \mathrm{d}\bar{\alpha}_{n}^{11} \right)^{x_{n-\hat{1}}^{11}} \left( \mathrm{d}\bar{\alpha}_{n}^{21} \right)^{x_{n-\hat{1}}^{21}} \left( \mathrm{d}\bar{\alpha}_{n}^{12} \right)^{x_{n-\hat{1}}^{12}} \left( \mathrm{d}\bar{\alpha}_{n}^{22} \right)^{x_{n-\hat{1}}^{22}} \nonumber \\
  & \cdot \left( \mathrm{d}\bar{\beta}_{n}^{11} \right)^{x_{n-\hat{2}}^{11}} \left( \mathrm{d}\bar{\beta}_{n}^{21} \right)^{x_{n-\hat{2}}^{21}} \left( \mathrm{d}\bar{\beta}_{n}^{12} \right)^{x_{n-\hat{2}}^{12}} \left( \mathrm{d}\bar{\beta}_{n}^{22} \right)^{x_{n-\hat{2}}^{22}} \nonumber \\
  & \cdot \left[ \prod_{a,b=1}^{2} \left( \bar{\alpha}_{n+\hat{1}}^{ab}\alpha_{n}^{ab} \right)^{x_{n}^{ab}} \left( \bar{\beta}_{n+\hat{2}}^{ab}\beta_{n}^{ab} \right)^{t_{n}^{ab}} \right].
\end{align}
Note that these tensor elements are quite similar to the tensor network representation of the Majorana--Wilson fermion system given in the authors' previous paper~\cite{Asaduzzaman:2022pnw}.
Indeed, if one takes a mapping as
$11 \to 1$,
$21 \to 2$,
$12 \to 3$,
and $22 \to 4$,
$G$ is exactly the same as that in~\cite{Asaduzzaman:2022pnw}.

The total partition function is then
\begin{align}
  Z = \sum_{\{x,t\}} \int \mathcal{D}U \prod_{n}
  \begin{aligned}[t]
    & T_{\mathrm{F}} G_{n}
      \left[ \prod_{a,b=1}^{2} \left( \mathcal{U}_{n,1}^{ab} \right)^{x_{n}^{ab}} \left( \mathcal{U}_{n,2}^{ab} \right)^{t_{n}^{ab}} \right]
      \left[ \prod_{a,b,c,d=1}^{2} e^{(\beta/2) U_{n,1}^{ab} U_{n+\hat{1},2}^{bc} U_{n+\hat{2},1}^{dc\ast} U_{n,2}^{ad\ast}} \right].
  \end{aligned}
\end{align}
Note that for the gauge part of the action  we can use the normal link variables $U$ rather than the projected ones $\mathcal{U}$ since the real part of $UUUU$ and $\mathcal{U}\mathcal{U}\mathcal{U}\mathcal{U}$ are the same.

To consider the integral of the gauge variables, we use the following
parametrization of the gauge elements

\begin{align}
  \label{eq:gaugeparametrization}
  U_{n,\mu}\left( \theta, \alpha, \gamma \right) =
  \begin{pmatrix}
    \cos\theta_{n,\mu} e^{i\alpha_{n,\mu}} & \sin\theta_{n,\mu}e^{i\gamma_{n,\mu}} \\
    -\sin\theta_{n,\mu} e^{-i\gamma_{n,\mu}} & \cos\theta_{n,\mu}e^{-i\alpha_{n,\mu}}
  \end{pmatrix}. 
\end{align}

Under this parametrization the Haar measure becomes
\begin{align}
  \int\mathcal{D}U = \int \prod_{n,\mu} \mathrm{d}U_{n,\mu}
  = \prod_{n,\mu} \int_{0}^{\frac{\pi}{2}} \mathrm{d}\theta_{n,\mu} \int_{-\pi}^{\pi} \mathrm{d}\alpha_{n,\mu} \int_{-\pi}^{\pi} \mathrm{d}\gamma_{n,\mu} \frac{\sin\theta_{n,\mu} \cos\theta_{n,\mu}}{2\pi^{2}}.
\end{align}

We can now discretize the variables by using the Gaussian quadrature rule.
For example, for a single variable function $g$, the Gauss--Legendre (GL) quadrature rule is
\begin{align}
  \int_{a}^{b} \mathrm{d}y \, g(y) \approx \frac{b-a}{2} \sum_{i=1}^{K} w_{i} g\left( \frac{b-a}{2}z_{i} + \frac{a+b}{2} \right).
\end{align}
$K$ is the order of the Legendre polynomial to be used, $z_i$ is the root of the Legendre polynomial, and $w_i$ is the corresponding weight. The higher the order $K$ of the polynomial is, the better the approximation of the integral is.
The formula generalizes to multi variable integrals 

\begin{align}
  \left(\prod_{i}\int_{a_i}^{b_i} \mathrm{d}y_i\right) \, g(y_1, \ldots, y_i, \ldots) \approx & \left(\prod_i \frac{b_i-a_i}{2} \right) \left(\prod_i \sum_{i=1}^{K}\right)  \left(\prod_i w_{i}\right) \nonumber \\ & g\left( \frac{b-a}{2}z_{1} + \frac{a_i+b_i}{2},\ldots, \frac{b_i-a_i}{2}z_{i} + \frac{a_i+b_i}{2},\ldots \right).
\end{align}
Using this discretization
each plaquette interaction factor can be regarded as a twelve rank tensor
\begin{align}
\label{eq:ptensor}
  P_{(ijk)(lmn)(opq)(rst)}
  & = \prod_{a,b,c,d=1}^{2} e^{(\beta/2)U^{bc}U^{dc\ast}U^{ad\ast}U^{ab}} \nonumber \\
  & =
  \begin{aligned}[t]
  \prod_{a,b,c,d=1}^{2} \exp\Biggl\{
    & \frac{\beta}{2}
      U\left(\frac{\pi}{4}z_{i} + \frac{\pi}{4}, \pi z_{j}, \pi z_{k}\right)_{bc}
        U\left(\frac{\pi}{4}z_{l} + \frac{\pi}{4}, \pi z_{m}, \pi z_{n}\right)_{dc}^{\ast} \nonumber \\
      & \cdot U\left(\frac{\pi}{4}z_{o} + \frac{\pi}{4}, \pi z_{p}, \pi z_{q}\right)_{ad}^{\ast}
        U\left(\frac{\pi}{4}z_{r} + \frac{\pi}{4}, \pi z_{s}, \pi z_{t}\right)_{ab}
    \Biggr\},
    \end{aligned}
\end{align}
where $z$-variable corresponds to each one of the three angles in the parameterization of the gauge group element in eq.~\ref{eq:gaugeparametrization}.
For simplicity we omit showing the indices for coordinates and directions here.

The number of elements of $P$, namely $K^{12}$, still grows rapidly along with $K$,
but one wants to have large $K$ to keep the accuracy of the GL quadrature approximation.
To address the large rank of the tensor, the Tucker decomposition can be used to express $P$ as a product of lower rank tensors.
In this paper we apply the higher order orthogonal iteration (HOOI) algorithm~\cite{de2000best} to the plaquette tensor~\footnote{
  One can of course apply the higher order singular value decomposition (HOSVD)~\cite{de2000multilinear} to $P$.
  However, the HOOI has an advantage in terms of both CPU and memory.
  It is expected that the HOOI reproduces the result of the HOSVD.
  Indeed, in the numerical section in this paper, we will show convergence of this algorithm for some cases.
}.

The HOOI algorithm proceeds as follows.
\begin{enumerate}
  \setcounter{enumi}{-1}
\item Input: an $N$-rank tensor $A$ whose bond dimension is $\chi$. Output: a core tensor $C$, whose bond dimension is $\chi^{\prime} < \chi$, and a set of unitary matrices $V$ whose dimension is $\chi^{\prime} \times \chi$, so that the tensor
  \begin{align}
    X_{I_{1}I_{2} \cdots I_{N}} = \sum_{i_{1},i_{2}, \ldots, i_{N}=1}^{\chi^{\prime}} C_{i_{1}i_{2} \cdots i_{N}} V^{[1]}_{i_{1}I_{1}} V^{[2]}_{i_{2}I_{2}} \cdots V^{[N]}_{i_{N}I_{N}}
  \end{align}
  approximates $A$ well.
  For the simplicity, here we assume that the length of each direction is the same for each $A$ and $C$.
\item Initialize $V$s as randomly generated unitary matrices.
\item For $j$-th leg each,
  \begin{itemize}
  \item Apply $V^{\left[ \tilde{j} \right]\dagger}$s to $A$ for $\tilde{j} \neq j$:
    \begin{align}
      B_{i_{1}i_{2} \cdots I_{j} \cdots i_{N}} = 
      \sum_{\substack{I_{1},I_{2},\ldots,I_{j-1},I_{j+1},\ldots,I_{N} =1}}^{\chi} 
      A_{I_{1}I_{2} \cdots I_{N}} V^{[1]\dagger}_{I_{1}i_{1}} V^{[2]\dagger}_{I_{2}i_{2}} \cdots V^{[j-1]\dagger}_{I_{j-1}i_{j-1}} V^{[j+1]\dagger}_{I_{j+1}i_{j+1}} \cdots V^{{[N]\dagger}}_{I_{N}i_{N}},
    \end{align}
  \item Take a truncated singular value decomposition (SVD) for the $j$-th leg of $B$:
    \begin{align}
      B_{i_{1}i_{2} \cdots I_{j} \cdots i_{N}} \approx \sum_{k=1}^{\chi^{\prime}} O_{i_{1}i_{2} \cdots k \cdots i_{N}} \rho_{k} P^{\dagger}_{kI_{j}},
    \end{align}
  \item Update $V^{[j]}$ by $P^{\dagger}$.
  \end{itemize}
\item Update $C$ as
  \begin{align}
    C_{i_{1}i_{2} \cdots i_{N}} = \sum_{I_{1},I_{2},\ldots,I_{N}=1}^{\chi} A_{I_{1}I_{2} \cdots I_{N}} V^{[1]\dagger}_{I_{1}i_{1}} V^{[2]\dagger}_{I_{2}i_{2}} \cdots V^{{[N]\dagger}}_{I_{N}i_{N}}.
  \end{align}
\item Iterate until the error $|A - X|_{\mathrm{F}}/|A|_{\mathrm{F}}$ converges, where $\left| \ \cdot \ \right|_{\mathrm{F}}$ denotes the Frobenius norm.
\end{enumerate}

HOOI has a quite tolerable numerical complexity to HOSVD, where SVDs are taken for each leg of $A$ directly.
Another big advantage of HOOI is that one does not need to store $P$ explicitly in memory.
Instead, one can just calculate an element of $P$ on demand.
Of course this is a tradeoff with computational complexity.

After applying the HOOI, the plaquette tensor $P$ is decomposed into a core tensor $S$ and a set of unitary matrices $V$:
\begin{align}
\label{eq:plaquettedecomp}
  & P_{\zeta_{n+\hat{1},2}\zeta_{n+\hat{2},1}\zeta_{n,2}\zeta_{n,1}} \nonumber \\ 
  \approx & \sum_{ \substack{x_{n,\mathrm{b}},t_{n,\mathrm{b}},x_{n-\hat{1},\mathrm{b}},t_{n-\hat{2},\mathrm{b}}=1} 
  }^{D}
  S_{x_{n,\mathrm{b}} t_{n,\mathrm{b}} x_{n-\hat{1},\mathrm{b}} t_{n-\hat{2},\mathrm{b}}} V^{[1]}_{x_{n,\mathrm{b}}\zeta_{n+\hat{1},2}} V^{[2]}_{t_{n,\mathrm{b}}\zeta_{n+\hat{2},1}} V^{[3]}_{x_{n-\hat{1},\mathrm{b}}\zeta_{n,2}} V^{[4]}_{t_{n-\hat{2},\mathrm{b}}\zeta_{n,1}},
\end{align}
where $D < K^{3}$ and where each $\zeta$ simply denotes a set of three indices that correspond to the roots of the Legendre polynomial (see eq.~\eqref{eq:ptensor} for the correspondence).
In this way one can approximate the plaquette tensor with a memory
requirement of $\mathcal{O}(D^{4} + 4DK^{3})$ instead of $\mathcal{O}(K^{12})$.

Finally, the full partition function is
\begin{align}
\label{eq:tnz}
  Z = \sum_{\{x,t\}} \prod_{n} \sum_{\zeta_{n,1}, \zeta_{n,2}, x_{n-\hat{1},\mathrm{b}}^{\prime}, t_{n-\hat{2},\mathrm{b}}^{\prime}}
  \begin{aligned}[t]
    & T_{\mathrm{F}} G_{n} S_{x_{n,\mathrm{b}} t_{n,\mathrm{b}} x_{n-\hat{1},\mathrm{b}}^{\prime} t_{n-\hat{2},\mathrm{b}}^{\prime}}
      \left[ \prod_{a,b=1}^{2} \mathcal{U}_{n,1}^{ab}\left( \zeta_{n,1} \right)^{x_{n}^{ab}} \mathcal{U}_{n,2}^{ab}\left( \zeta_{n,2} \right)^{t_{n}^{ab}} \right] \\
    & \cdot 
    V^{[4]}_{t_{n-\hat{2},\mathrm{b}}^{\prime} \zeta_{n,1}}
    V^{[2]}_{t_{n-\hat{2},\mathrm{b}}\zeta_{n,1}}
    V^{[3]}_{x_{n-\hat{1},\mathrm{b}}^{\prime} \zeta_{n,2}} 
    V^{[1]}_{x_{n-\hat{1},\mathrm{b}} \zeta_{n,2}},
  \end{aligned}
\end{align}
where the summation for $\zeta_{n,1}$, $\zeta_{n,2}$ and for $x_{n-\hat{1},\mathrm{b}}^{\prime}$, $t_{n-\hat{2},\mathrm{b}}^{\prime}$ run over $K^{3}$ and $D$ integers, respectively~\footnote{
Note that we assume the weight and the constant factors generated from the Gaussian quadrature are incorporated to $P$ tensor.
Otherwise one should explicitly have the factors in eq.~\eqref{eq:tnz}.
}.
By defining the integrated bosonic tensor as
\begin{align}
  T_{x_{n} t_{n} x_{n-\hat{1}} t_{n-\hat{2}}} =
  \sum_{\zeta_{n,1}, \zeta_{n,2}, x_{n-\hat{1},\mathrm{b}}^{\prime}, t_{n-\hat{2},\mathrm{b}}^{\prime}}
  & T_{\mathrm{F} x_{n,\mathrm{f}} t_{n,\mathrm{f}} x_{n-\hat{1},\mathrm{f}} t_{n-\hat{2},\mathrm{f}}} S_{x_{n,\mathrm{b}} t_{n,\mathrm{b}} x_{n-\hat{1},\mathrm{b}}^{\prime} t_{n-\hat{2},\mathrm{b}}^{\prime}}
    \left[ \prod_{a,b=1}^{2} \mathcal{U}_{n,1}^{ab}\left( \zeta_{n,1} \right)^{x_{n}^{ab}} \mathcal{U}_{n,2}^{ab}\left( \zeta_{n,2} \right)^{t_{n}^{ab}} \right] \nonumber \\
  &\cdot 
  V^{[4]}_{t_{n-\hat{2},\mathrm{b}^{\prime}} \zeta_{n,1}} 
  V^{[2]}_{t_{n-\hat{2},\mathrm{b}}\zeta_{n,1}}
  V^{[3]}_{x_{n-\hat{1},\mathrm{b}^{\prime}} \zeta_{n,2}} 
  V^{[1]}_{x_{n-\hat{1},\mathrm{b}} \zeta_{n,2}},
\end{align}
the partition function can be written as
\begin{align}
  Z = \sum_{\{x,t\}} \prod_{n}
  T_{x_{n} t_{n} x_{n-\hat{1}} t_{n-\hat{2}}} G_{n, x_{n} t_{n} x_{n-\hat{1}} t_{n-\hat{2}}}.
\end{align}
In this expression the indices with the subscript ``$\mathrm{f}$'' denote the set of fermionic (binary) indices; \textit{i.e.} $x_{\mathrm{f}} = (x^{11},x^{21},x^{12},x^{22})$.
Also, the integrated indices are simply shown without subscript as in $x = (x_{\mathrm{f}},x_{\mathrm{b}})$.

\section{Numerical results}

\subsection{Pure $SU(2)$ gauge theory}

\begin{figure}[htbp]
  \centering
  \includegraphics[width=\hsize]{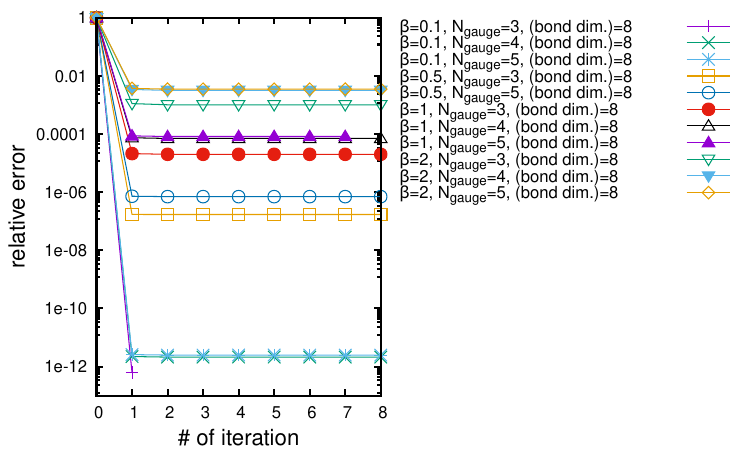}
  \caption{Relative errors of the HOOI for the $SU(2)$ plaquette tensor.}
  \label{fig:errhooi_bdim8}
\end{figure}

Figure~\ref{fig:errhooi_bdim8} shows how the relative error converges for the $SU(2)$ plaquette tensor as the HOOI proceeds.
Here we discretize the plaquette tensor by using the roots of the Legendre polynomial with varying the number of roots $N_{\mathrm{gauge}}$ to be $3$, $4$, and $5$.
$N_{\rm gauge}$ in this section is identified as $K$ in the previous section; in other words, we approximate the plaquette tensor by replacing the integrals of angle by summations over the $N_{\mathrm{gauge}}$ roots of the Legendre polynomial.
With the same notation in eq.~\eqref{eq:plaquettedecomp}, the error in the figure is defined by
\begin{align}
    \frac{\left| P - SV^{[1]}V^{[2]}V^{[3]}V^{[4]} \right|_{\mathrm{F}}}{\left| P \right|_{\mathrm{F}}}.
\end{align}
From the figure we can observe that larger $\beta$ are relatively difficult although
fortunately the iteration rapidly converges in all cases.
Surprisingly, in the strong coupling region $\beta < 0.5$, the accuracies are beyond the single precision even though the drastic reduction of the number of d.o.f. (from $N_{\mathrm{gauge}}^{3\times 4}$ to $8^{4}$) is taken place.

\begin{figure}[htbp]
  \centering
  \includegraphics[width=0.8\hsize]{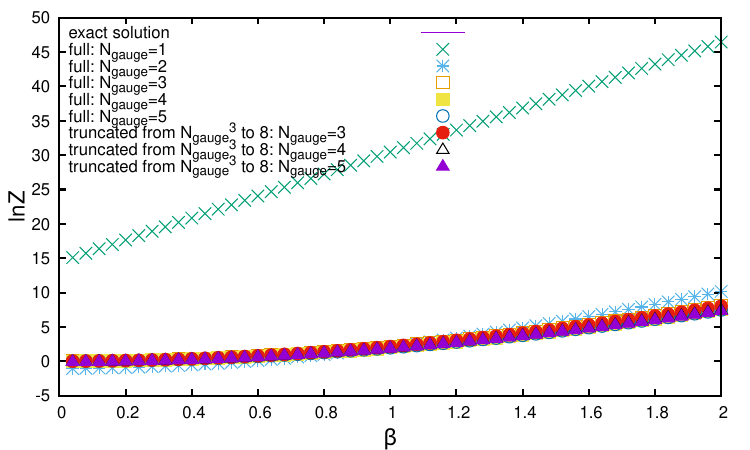}
  \caption{Free energy for pure $SU(2)$ gauge theory as a function of $\beta$ in two
    dimensions.
    Exact solution is reproduced with 2--3 digits accuracy, so that the exact (purple) line is behind the data points.
    }
  \label{fig:F_vs_beta_pure}
\end{figure}

\begin{figure}[htbp]
  \centering
  \includegraphics[width=0.8\hsize]{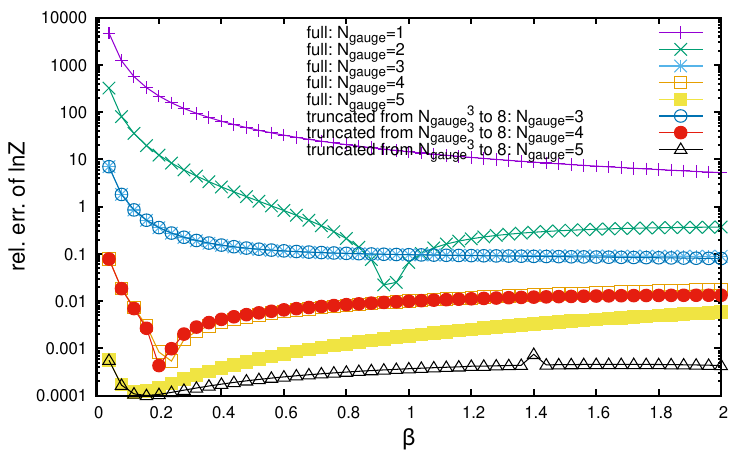}
  \caption{Relative error in the free energy for pure $SU(2)$ in two
    dimensions.}
  \label{fig:Rel_error_free_pure}
\end{figure}

Next we show the efficiency of the truncated quadrature scheme by comparing free energies calculated from the tensor renormalization group with the exact solution.
The latter is easy to derive in two dimensions since the partition function can be reduced to a single plaquette integral. 
For the sake of completeness, the partition function of the pure $SU(2)$ gauge theory in terms of tensors is detailed in the appendix~\ref{appendix_pure_su2} using the character expansion.
Figures \ref{fig:F_vs_beta_pure} and \ref{fig:Rel_error_free_pure} show the free energy of the pure $SU(2)$ theory and corresponding relative errors on a $L=4$ lattice. In these figures,
``Full'' indicates that the plaquette tensor with $N_{\mathrm{gauge}}^{3 \times 4}$ elements is treated as the fundamental tensor in the network.
On the other hand, truncated cases are also shown, where $N_{\mathrm{gauge}}^{3 \times 4}$ elements are reduced to $8^{4}$ (fixed for any choice of $N_{\mathrm{gauge}}$) by using the HOOI algorithm.
It is clear from the error analysis that relatively a small number of terms (\textit{i.e.} $N_{\mathrm{gauge}}$) is needed in the quadrature approximation
and that the effect of the further reduction by the HOOI is quite small.

We also find from the comparison to the relative errors in fig.~\ref{fig:errhooi_bdim8} that the $\beta$ dependence is quite milder for the free energies.
This might be attributed to some cancellation occurring among neighboring plaquettes.

\subsection{$SU(2)$ theory coupled to reduced staggered fermions}

\begin{figure}[htbp]
  \centering
  \includegraphics[width=0.8\hsize]{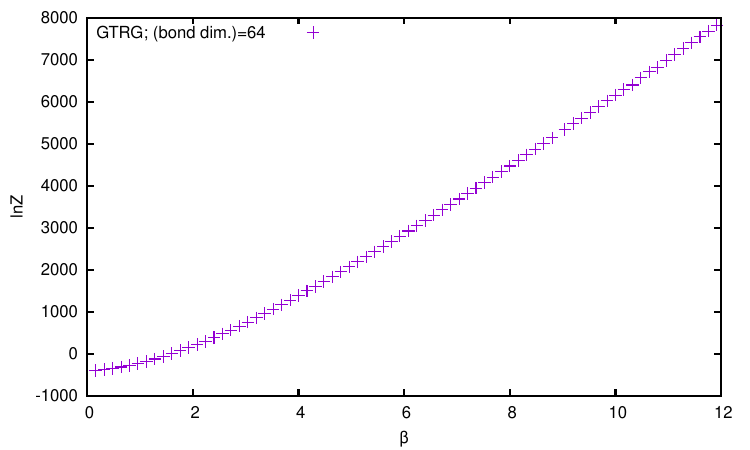}
  \caption{Free energy of (massless) reduced staggered fermions coupled to $SU(2)$ gauge
  fields on $32\times 32$ lattice.
  }
  \label{F_vs_beta_full}
\end{figure}

We now turn to the theory including reduced staggered fermions. Figure~\ref{F_vs_beta_full} shows
a plot of the free energy versus $\beta$ on $L=32$ lattice.

\begin{figure}[htbp]
  \centering
  \includegraphics[width=0.8\hsize]{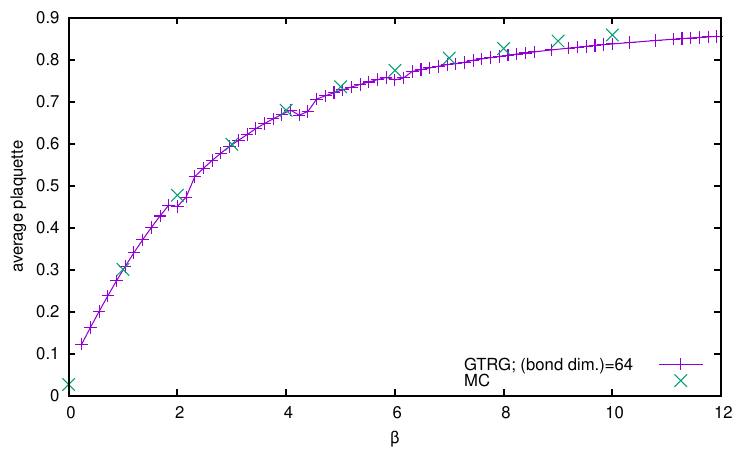}
  \caption{Average plaquette for reduced staggered fermions coupled to $SU(2)$ on $L=32$ lattice.}
  \label{plaq_l32}
\end{figure}

To check for the accuracy of the tensor network calculation we have compared the expectation
value of the plaquette with Monte Carlo results~\footnote{In general the Pfaffian
  arising in reduced staggered fermions suffers from a sign problem, but one can use
  the pseudoreal property of the gauge group
  to show that this is evaded in the case of $SU(2)$. It can hence be simulated
  with a conventional RHMC algortithm.}. This comparison is shown in fig.~\ref{plaq_l32}
for a lattice of size $L=32$ and a bond dimension of $64$.

\begin{figure}[htbp]
  \centering
  \includegraphics[width=0.8\hsize]{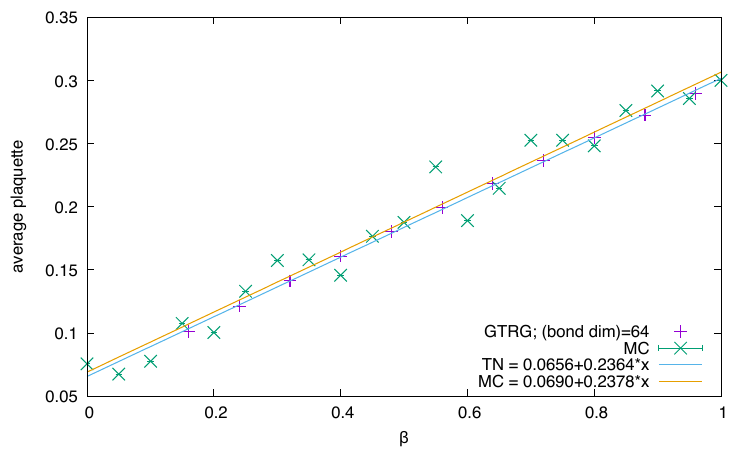}
  \caption{
  Plaquette on $L=8$ lattice in the small $\beta$ region compared with strong coupling.}
  \label{small_beta}
\end{figure}

Clearly the Monte Carlo agrees well with the tensor network result over a wide
range of $\beta$. It is interesting to examine in more detail the small $\beta$ region. This
is done in fig.~\ref{small_beta}. The straight line shows a fit to the strong
coupling result for the average plaquette $P$
\begin{align}
  P=\left(\frac{1}{2}\right)^4+\frac{\beta}{4},
  \end{align}
where the intercept arises from the leading contribution to the plaquette from expanding the fermion hopping term.
One can see the stability of TN result and that the TN calculation finely reproduces the analytical formula.

\section{Summary}

In this paper we have shown how to construct a tensor network representing the path integral
of reduced staggered fermions coupled to an $SU(2)$ gauge field which is minimal
in terms of its memory and computational requirements. We have described
the complexities arising in formulating tensor network representations for fermions coupled
to non-abelian gauge fields and shown how the use of reduced staggered fermions combined with
a HOOI modified Gaussian quadrature algorithm for handling the gauge fields,
allows for an efficient tensor representation. We use this representation to compute the free
energy and the average plaquette using the Grassmannn tensor renormalization group
(GTRG) algorithm  finding good agreement with Monte Carlo results in the case of the
latter.

In general one expects that $SU(N)$ gauge theories coupled to reduced staggered fermions
will have sign problems and this is hence the arena in which tensor formulations such as the
one described in this paper will be most useful. We hope to report on such work in
the near future.

\acknowledgments
We thank the members of the QuLAT Collaboration for valuable discussions.
This work was supported in part by the U.S. Department of Energy (DOE) under Award Numbers DE-SC0009998, DE-SC0010113, and DE-SC0019139.
This research used resources of
the Syracuse University HTC Campus Grid and NSF award ACI-1341006
and
the National Energy Research Scientific Computing Center (NERSC), a U.S. Department of Energy Office of Science User Facility located at Lawrence Berkeley National Laboratory, operated under Contract No. DE-AC02-05CH11231 using NERSC awards HEP-ERCAP0020659 and HEP-ERCAP0023235.

%

\appendix

\section{Character expansion formulae}

The character expansion is given by
\begin{align}
  \label{eq:charexp}
  e^{\left( \beta/2 \right) \tr \left[ U_{n,1} U_{n+\hat{1},2} U_{n+\hat{2},1}^{\dagger} U_{n,2}^{\dagger} \right]}
  = \sum_{r_{n}=0}^{\infty} F_{r_{n}}\left( \beta \right) \chi_{r_{n}}\left( U_{n,1} U_{n+\hat{1},2} U_{n+\hat{2},1}^{\dagger} U_{n,2}^{\dagger} \right).
\end{align}
For the $SU(2)$ case, $F$ is expressed using the modified Bessel function of the first kind $I$:
\begin{align}
  \label{eq:charexpcoeff}
  F_{r}\left( \beta \right) = I_{2r}\left( \beta \right) - I_{2r+2}\left( \beta \right)
  = 2 \left( 2r+1 \right) \frac{I_{2r+1}\left( \beta \right)}{\beta}.
\end{align}
$\chi$ is called the character, whose properties are given below.

The character of the product of the group elements can be broken up into the trace over the product of the matrix represnetation of the group elements:
\begin{align}
  \label{eq:charbreakup}
  \chi_{r_{n}}\left( U_{n,1} U_{n+\hat{1},2} U_{n+\hat{2},1}^{\dagger} U_{n,2}^{\dagger} \right)
  = \sum_{a,b,c,d} D^{[r_{n}]}_{ab}\left( U_{n,1} \right) D^{[r_{n}]}_{bc}\left( U_{n+\hat{1},2} \right) D^{[r_{n}]\dagger}_{cd}\left( U_{n+\hat{2},1} \right) D^{[r_{n}]\dagger}_{da}\left( U_{n,2} \right)
\end{align}
Note that the dimensions of the matrices (the ranges of $a$, $b$, $c$, $d$) depend on the label of the irreducible representation of the group $r$.
$D$ is called the Wigner D-matrix.

The D-matrices satisfy an orthogonality condition
\begin{align}
  \label{eq:dmatorthogonality}
  \int \mathrm{d}U D^{[r_{1}]}_{i_{1}j_{1}}\left( U \right) D^{[r_{2}]\ast}_{i_{2}j_{2}}\left( U \right)
  = \frac{1}{2r_{1}+1} \delta_{r_{1}r_{2}} \delta_{i_{1}i_{2}} \delta_{j_{1}j_{2}}.
\end{align}

\section{Pure $SU(2)$ with character expansion \label{appendix_pure_su2}}

The lattice action of the 2D pure $SU(2)$ model is given by
\begin{align}
  S = - \frac{\beta}{2} \sum_{n} \tr \left[ U_{n,1} U_{n+\hat{1},2} U_{n+\hat{2},1}^{\dagger} U_{n,2}^{\dagger} \right]
\end{align}
with the inverse coupling constant $\beta=1/g^{2}$ and the link variables $U_{n,\mu}=\exp\{igA^{i}_{n,\mu}T^{i}\}$.
$T$ is the generator of $SU(2)$.

We make a tensor network representation of the partition function
\begin{align}
  Z &= \int \mathcal{D}U e^{-S} \nonumber \\
    &= \int \mathcal{D}U \prod_{n} e^{\left( \beta/2 \right) \tr \left[ U_{n,1} U_{n+\hat{1},2} U_{n+\hat{2},1}^{\dagger} U_{n,2}^{\dagger} \right]},
\end{align}
where $\mathcal{D}U = \prod_{n} \mathrm{d}U_{n,1} \mathrm{d}U_{n,2}$ is the $SU(2)$ Haar measure.
By using the well known formulae~\eqref{eq:charexp},~\eqref{eq:charbreakup}, the partition function can be written using the Wigner D-matrices:
\begin{align}
  Z &= \sum_{\left\{ r,x,t,x^{\prime},t^{\prime} \right\}}
      \prod_{n}
      \begin{aligned}[t]
        & F_{r_{n}}\left( \beta \right)
          \int \mathrm{d}U_{n,1}
          D^{[r_{n}]}_{t^{\prime}_{n,1} t^{\prime}_{n,2}}\left( U_{n,1} \right)
          D^{[r_{n-\hat{2}}]\ast}_{t_{n-\hat{2},1} t_{n-\hat{2},2}}\left( U_{n,1} \right) \\
        & \cdot \int \mathrm{d}U_{n,2}
          D^{[r_{n}]\ast}_{x^{\prime}_{n,1} x^{\prime}_{n,2}}\left( U_{n,2} \right)
          D^{[r_{n-\hat{1}}]}_{x_{n-\hat{1},1} x_{n-\hat{1},2}}\left( U_{n,2} \right) \\
        & \cdot \delta_{t^{\prime}_{n,2} x_{n,1}} \delta_{x_{n,2} t_{n,2}} \delta_{t_{n,1} x^{\prime}_{n,2}} \delta_{x^{\prime}_{n,1} t^{\prime}_{n,1}}.
      \end{aligned}
\end{align}
The summation $\sum_{\left\{ \cdot \right\}}$ denotes the summation over the corresponding indices all over the sites and links; this rule is inherited throughout this paper.

Now we can integrate out the original link variables by using the orthogonality condition~\eqref{eq:dmatorthogonality} and obtain a tensor network representation:
\begin{align}
  Z &= \sum_{\left\{ r,x,t \right\}}
      \prod_{n}
      \frac{F_{r_{n}}\left( \beta \right)}{\left( 2r_{n}+1 \right)^{2}}
      \delta_{r_{n} r_{n-\hat{1}}}
      \delta_{r_{n} r_{n-\hat{2}}}
      \delta_{t_{n-\hat{2},2} x_{n,1}} \delta_{x_{n,2} t_{n,2}} \delta_{t_{n,1} x_{n-\hat{1},2}} \delta_{x_{n-\hat{1},1} t_{n-\hat{2},1}}.
      \label{eq:puresu(2)partfunc}
\end{align}
An object in the product in the righthand side can be regarded as a tensor placed on the center of each plaquette.

Note that all the indices associated to plaquette ($r$ in eq.~\eqref{eq:puresu(2)partfunc}) take the same value in two dimensions.
In other words, if one fixes one $r$, every other $r$ takes the same value owing to $\delta_{r_{n} r_{n-\hat{1}}}$ and $\delta_{r_{n} r_{n-\hat{2}}}$.
One may call this property the Gauss's law.

\end{document}